# Resolving the structural stability of the hydrogen storage borohydride $NaCa(BH_4)_3$ via data-driven screening and first-principles lattice dynamics

Sanaa Ismail[1,2,*], Ricardo Amaral[3], Attia A. Gadallah[4], Hassan Mohamed El-Said Azzazy[5], and Zi-Kui Liu[1]

[1] Department of Materials Science and Engineering, The Pennsylvania State University, University Park, PA 16802, USA

[2] Nanotechnology Program, School of Sciences and Engineering, The American University in Cairo, AUC Avenue, P.O. Box 74, New Cairo 11835, Egypt

[3] Department of Energy and Mineral Engineering, The Pennsylvania State University, University Park, PA 16802, USA

[4] Independent Researcher, State College, 16801 PA, USA

[5] Department of Biological, Chemical, Global Health Sciences, School of Sciences and Engineering, The American University in Cairo, AUC Avenue, P.O. Box 74, New Cairo 11835, Egypt

*Corresponding author: sii5085@psu.edu

## Abstract

Equimolar $NaCa(BH_4)_3$ offers a theoretical gravimetric **hydrogen capacity** of **11.24** wt.%, yet attempts to synthesize it within perovskite frameworks have been unsuccessful, and its crystal structure remains undetermined. Every structural model proposed for it derives from the **perovskite family** of its heavier homologs. Here, we search for its ground state without the perovskite restriction. The search combines 2238 database-derived candidates, exhaustive Na/Ca cation ordering at fixed stoichiometry, screening with a machine-learning interatomic potential, dispersion-corrected density functional theory, and lattice-dynamics verification. The ground-state structure identified is not a perovskite but **a monoclinic phase** of space group **Cm**, energetically favored over the most stable perovskite-type candidate by 7.67 meV/atom. Two unrelated intermetallic donors recover it independently, and its ordered cation arrangement is the lowest of the series, so a search over chemically intuitive structure types would have missed it. Against $NaBH_4$ and $Ca(BH_4)_2$, which bound the Na-Ca-B-H hull at this composition, the phase is metastable by only 3.37 meV/atom. That margin falls under a quarter of the median for observed inorganic crystals and places the release enthalpy 1.1 kJ per mole of $H_2$ below that of the precursors. Phonon calculations reveal no imaginary modes at the commensurate wavevectors, and the positive-definite elastic tensor establishes mechanical stability. The competing perovskite-type candidates carry imaginary modes and are therefore not local minima. The failure to synthesize this compound therefore stems from thermodynamic metastability against the binary precursors rather than from dynamical instability, and five X-ray reflections identify the phase for non-equilibrium synthesis attempts.

## 1 Introduction

The transition to zero-emission energy infrastructure depends on storing surplus renewable power and on reducing exposure to fossil-fuel markets [1,2]. Hydrogen is attractive in that role for its gravimetric energy density of 120 MJ $kg^{-1}$ [3,4], but its low volumetric density as a gas or a liquid makes storage and transport the limiting step [5,6], which has directed attention to solid-state storage. Light-metal borohydrides are among the highest-capacity candidates, with $NaBH_4$ holding 10.6 wt.% hydrogen and $Ca(BH_4)_2$ 11.6 wt.% [7–9]. However, neither is practical on its own, because desorption is slow and the dehydrogenation temperature is high [10]. Heterovalent bimetallic substitution places a monovalent alkali and a divalent alkaline-earth cation on the same borohydride lattice. The thermodynamics of $M(BH_4)_n$ **correlate with the** electronegativity of its cation [11–14], **so substitution tunes the dehydrogenation temperature.** The equimolar composition $NaCa(BH_4)_3$ carries a theoretical capacity of 11.24 wt.%, between those of its two parents. High-throughput screening established the promise of mixed alkali/alkaline-earth borohydride chemistries [15], and later first-principles work on the Ca-M (M = Li, Na, K)-B-H systems predicted that combining $NaBH_4$ with $Ca(BH_4)_2$ yields destabilized, reversible hydrogen-storage reactions [16]. Those studies worked from a small number of assumed structures, however, and carried out neither the exhaustive cation-ordering enumeration nor the lattice-dynamics test that identifying a structure requires. Machine learning has since been applied across solid-state hydrogen storage as regression of a property onto composition and structure [17–19], and to hydride perovskites in particular [20]. In each case the structure on which the property is predicted is assumed rather than obtained from a search.

No experimentally determined crystal structure of $NaCa(BH_4)_3$ is available in the literature, so its most stable form must be predicted rather than measured, and the result can be no more reliable than the candidate set the search is carried out on [21,22]. Two sources of error follow from that dependence. The first, selection bias, arises because a candidate set assembled from chemically intuitive structure types cannot demonstrate that the global minimum has been located, since the search never leaves the topology those types share [23]. Second, resolution arises because the energies separating competing lattice structures and cation arrangements in this system lie within a few meV/atom, below the accuracy reported for the machine-learning potentials used to screen them [24–26]. A ranking from such a potential alone cannot be treated as converged. A third requirement is independent of both, since a static minimum is a structure only if it is also a minimum with respect to atomic displacement. No structure search for $NaCa(BH_4)_3$ has gone beyond the perovskite analogs, and none of the structures proposed for it has been tested against its own lattice dynamics. We therefore searched the composition without perovskite restriction, enumerated the cation arrangement at fixed equimolar stoichiometry, screened the candidates with a machine-learning potential, and relaxed the survivors from first principles before computing their phonon spectra. The lowest-energy structure is monoclinic, space group Cm, and is not a perovskite. It lies 7.67 meV/atom lower in energy than the best perovskite-type arrangement and has no imaginary mode at any wavevector its supercell resolves exactly, whereas the perovskite-derived candidates retain soft modes. Together these steps form a workflow (Figure 1) in which large-scale structural exploration is followed by first-principles validation and a lattice-dynamics test, so that every candidate passing one stage is checked by a method the earlier stage cannot replace. The ground state it identifies derives from intermetallic donor prototypes, and each of the four structure types considered before this search lies higher in energy. Because the workflow needs no measured structure, it applies to other complex hydrides for which none exists.

## 2 Methods

Figure 1 outlines the workflow in four stages and its outcome. Candidate lattice structures come from experimentally determined analogs and from a database-wide search that is not restricted to any structure type (Figure 1, stages 1 and 2, Section 2.1). Every candidate is relaxed with a machine-learning interatomic potential (Figure 1, stage 1), and the Na/Ca arrangement on each retained lattice structure is enumerated exhaustively and checked against an independent enumeration on the native cells (Figure 1, stage 3, Section 2.2). The leading configurations are relaxed with dispersion-corrected density functional theory, which fixes the energy ranking and the stability against decomposition (Figure 1, stages 1 and 3, Section 2.3). Phonon and elastic calculations then test whether each static minimum is dynamically and mechanically stable, and simulated powder patterns give the reflections by which the predicted phase can be recognized (Figure 1, stage 4 and final panel, Section 2.4). Each stage narrows the candidate set or tests a property that the stage before it cannot resolve.

## 2.1 Candidate lattice structures and database-wide search

Six candidate lattice structures were assembled from three independent sources. Three are experimentally determined $ABX_3$ borohydrides of the same stoichiometry with the Na/Ca pair substituted: a cubic lattice structure (A, 17 atoms) [27], the $CsSr(BH_4)_3$ type (C, 34 atoms) and the $RbSr(BH_4)_3$ type (D, 68 atoms) [28]. A fourth, a triclinic distorted perovskite (B, 34 atoms), was obtained with crystALL from the GNoME entry a9b189292f [29]. B has the lowest energy among the perovskite-type arrangements the search returned, and it does not appear among the experimentally characterized $ABX_3$ borohydrides. Two further lattice structures, G1 and G2, came from the database-wide search. Table 1 lists their provenance and relaxed structures.

*crystALL* retrieved every *MPDD* entry whose anonymized formula matches $ABC_3D_{12}$ and performed ordered substitution of Na, Ca, B and H, returning 1119 parent prototypes, each generated in its two symmetry-distinct Na/Ca permutations for 2238 candidates. Because anonymized substitution transfers topology rather than chemistry (Supplementary Section S1), candidates were ranked only after relaxation by MACE. All 2238 were relaxed under the protocol of Section 2.2, none excluded a priori. Relaxed structures were deduplicated using pymatgen's *StructureMatcher* algorithm [30] under primitive-cell reduction and scaling. The five lowest in energy went to first-principles validation, where they reduced to three distinct lattice structures.

## 2.2 Cation-ordering enumeration and machine-learning relaxation

The cation arrangement is the only degree of freedom at the charge-neutral equimolar composition, in which one $Na^+$ and one $Ca^{2+}$ balance three $[BH_4]^-$ anions. The arrangements enumerated here are derivative structures of the parent lattice, generated by ordered substitution on the cation sublattice with the $[BH_4]^-$ tetrahedra held rigid [31], so that the enumeration acts on the cation sublattice alone (Supplementary Section S1). Two complementary enumerations were performed. The first served to validate the completeness of the configurational space across varying cell sizes. Lattice structure A, with only two cation sites per primitive cell, was enumerated with the Hart-Forcade algorithm [32,33] as implemented in ICET (release 3.2) [34] over all symmetry-distinct supercells up to index 4 (17 to 68 atoms) at fixed equimolar occupancy [35]. Lattice structures B, C, and D, with 4, 4, and 8 native cation sites, were enumerated over all $C(N, N/2)$ configurations. Across these four initial lattice structures, 2836 raw configurations reduce to 135 inequivalent configurations by orbit counting under the parent symmetry (Table S1).

The second enumeration generated the structures actually carried forward for the energy rankings. To allow a direct energetic comparison across the order parameter, this enumeration standardizes every lattice structure, including G1 and G2 from the database search, to exactly eight cation sites (A, G1 and G2 expanded to 1 × 2 × 2; B and C doubled along their shortest lattice vectors; D evaluated natively). The arrangement was then varied along a swap fraction $s = m/(n/2)$, where $m$ is the number of Na↔Ca exchanges from the ordered parent [36]. This standardization yields an identical raw space of $C(8,4) = 70$ configurations per lattice structure. Across all six frameworks, these 420 raw configurations reduce to 118 unique, symmetry-inequivalent arrangements over $s = 0, 0.25, 0.50, 0.75$, and 1 (Table 2). These 118 configurations were the ones passed to the machine-learning screen.

Because the orientational degrees of freedom of $[BH_4]^-$ lower the effective symmetry below the nominal space group, equivalence was judged from sorted nearest-neighbor distance fingerprints [37,38] rather than parent point-group operations. Completeness was verified against an independent ATAT enumeration [39–41], which returns identical native-cell counts, and by the identity count(s) = count(1−s).

All configurations were relaxed with the MACE-MP-0-medium potential [42,43] in 64-bit precision and without dispersion correction, using ASE [44], with a *FrechetCellFilter* to a force tolerance of 0.05 eV/Å. Database-search candidates were relaxed in two stages, isotropic strain at fixed coordinates and cell shape followed by full relaxation, so that over-expanded templates reach equilibrium volume without losing $[BH_4]^-$ integrity.

## 2.3 First-principles calculations

Calculations used VASP [45–47] within the projector augmented-wave formalism [48,49], with the PBE functional [50] and Grimme D3(BJ) dispersion (IVDW = 12) [51,52]. Sodium and calcium were described with the *Na_pv* and *Ca_pv* datasets, boron and hydrogen with the standard B and H datasets. The calculations used a 600 eV plane-wave cutoff, Γ-centered Brillouin-zone sampling at a uniform k-point spacing of 0.15 $Å^{-1}$ [53], Gaussian smearing ($\sigma$ = 0.05 eV), electronic convergence to $10^{-8}$ eV, residual forces below 0.01 eV/Å, and simultaneous relaxation of coordinates, volume and cell shape (ISIF = 3), so all reported energies lie on one scale. Because a fixed spacing rather than fixed divisions sets the k-point mesh, the grid dimensions follow the cell. They give 11 × 8 × 7 (309 irreducible points) for the 17-atom primitive cell of G2, 5 × 11 × 7 (193) for its 68-atom eight-cation-site cell, 10 × 10 × 8 (105) for NaBH4 and 9 × 9 × 9 (125) for Ca(BH4)2. The Na, Ca, B and $H_2$ references use 14 × 14 × 7, 8 × 8 × 8, 10 × 10 × 4 and 6 × 6 × 6. The formation energy per atom is

$$\Delta E_{form} = \frac{E_{DFT}(\mathrm{Na}_a Ca_b B_c H_d) - (aE(Na) + bE(Ca) + cE(B) + \frac{d}{2}E(H_2))}{a+b+c+d} \qquad \textbf{\textit{Eq. 1}}$$

where $E_{DFT}$ is the relaxed-cell energy and a to d the numbers of Na, Ca, B and H atoms. The elemental references were relaxed at the same settings, giving μ(Na) = −1.473, μ(Ca) = −2.156, μ(B) = −6.913 and μ(H) = −3.408 eV/atom (Table S2). Stability against the binary borohydrides was referenced to $NaBH_4$ in $P4_2/nmc$ (No. 137), Materials Project [54] mp-38725, at −24.082 eV per formula unit and to $Ca(BH_4)_2$ in C2/c (No. 15) at −47.132 eV per formula unit, together −4.189013 eV/atom. The C2/c structure was obtained by following the soft mode of Fddd polymorph of $Ca(BH_4)_2$ which is dynamically unstable. C2/c carries no imaginary mode (Supplementary Section S2). Space groups were assigned using the *spglib* Python library [55] with a tolerance of 0.05 Å.

## 2.4 Lattice dynamics and diffraction simulation

Phonon dispersions and densities of states were computed by finite displacements with *phonopy* [56,57], interfaced to VASP through *hydrophonokit* [58], for G2 and for B and C at s = 0.50, each re-relaxed to $10^{-3}$ eV/Å before displacements were generated. G2 used a 3 × 2 × 2 supercell (204 atoms, 102 symmetry-reduced displacements), sized on perpendicular cell widths of 12.1, 10.7 and 12.7 Å; B and C used 2 × 2 × 2 supercells (272 atoms, 204 and 51 displacements). Forces were converged to $10^{-8}$ eV with reciprocal-space projection (LREAL = .FALSE.) and ADDGRID = .TRUE. Born effective charges and dielectric tensors from DFPT [59,60] on the undisplaced primitive cells were applied as a non-analytical correction for LO-TO splitting, and dispersions mapped along symmetry-enforced paths [61]. The relaxed-ion elastic tensor of G2 was computed by finite strain (IBRION = 6) [62]. The machine-learning screen and the convergence tests are reported in Supplementary Section S2. The same protocol was applied to the two reference phases, $NaBH_4$ ($P4_2/nmc$, 3 × 3 × 2 supercell) and $Ca(BH_4)_2$ (C2/c, 2 × 2 × 2), neither of which carries an imaginary mode; their free energies enter the reaction free energy of Section 3.2.

Powder patterns were computed from the relaxed cells for Cu Kα radiation with the $K\alpha_1/K\alpha_2$ doublet resolved (1.540598 and 1.544426 Å, intensity ratio 2:1) and a pseudo-Voigt instrumental profile of width $FWHM^2 = U \tan^2\theta + V \tan\theta + W$ with U = 0.020, V = −0.002 and W = 0.008 deg², giving 0.090° at 13° in 2θ and 0.116° at 60°. Debye-Waller factors were taken from the phonon calculation rather than assumed. $B_{iso} = 8\pi^2\langle u^2\rangle$ at 0 K, with modes below 0.1 THz excluded, gives 0.27, 0.51, 0.51 and 2.09 Å² for Ca, B, Na and H, computed for G2 and applied to every phase. The zero-Kelvin value is used because the harmonic $\langle u^2 \rangle$ cannot represent the thermally activated $[BH_4]^-$ reorientation that sets in at measurement temperatures [63].

## 3 Results and discussion

### 3.1 The candidate lattice structures, cation ordering and the machine-learning screen

First-principles relaxation reduces Candidates 1-5 to three structures: C1 and C5 relax to G1, C2 and C4 to G2, and C3 to G3 (Table 3). Each pair comes from a different donor prototype and a different descriptor cluster of the crystALL map (Figure 2), and the two members of a pair are distinct structure types on the machine-learning surface. After relaxation they agree to root-mean-square displacements of 0.009 Å (G1) and 0.006 Å (G2) and to 0.017 and 0.076 meV/atom in energy. G1 and G2, the two lowest in energy of the three structures reached from C1 to C5, both enter the cation-ordering enumeration. Figure 3 shows the distribution of the candidate library across the prototype embedding and the energetic ranking of the hundred lowest-energy structures. All 2238 candidates were relaxed under a single protocol with none withheld from calculation, so the relaxation decides the outcome for each template rather than a choice made in advance.

Structurally, G2 occupies a 17-atom cell and stands as the most stable non-perovskite structure, while B occupies a 34-atom cell and stands as the most stable perovskite-type structure. The two differ in symmetry. A tolerance scan assigns G2 to the monoclinic space group Cm across the whole range 0.01 to 0.5 Å with no tolerance dependence, whereas B is triclinic throughout. The lattice structure recovered by the search is therefore not a further instance of the low-symmetry character already established for this system but a distinctly more ordered one, and it lies 7.67 meV/atom below B at s = 0.50.

From the standardized 8-site swap-fraction enumeration, the 420 initial configurations reduce to 118 unique representatives overall. The cubic prototype collapses to a single representative per swap fraction, while the triclinically distorted lattice retains the largest configurational diversity. Because all lattice structures originate from the same combinatorial space by construction (70 equimolar configurations of the eight-site cell, distributed 1/16/36/16/1 over s), the symmetry-reduced totals in Table 2 isolate the effect of lattice symmetry alone. The counts agree exactly with an independent ATAT enumeration on the native cells, 6, 4 and 22 for B, C and D. For the cubic lattice structure, a Hart-Forcade reference resolved by cell size gives 2, 4 and 83 configurations on 17-, 34- and 68-atom cells. The enumeration and its checks are documented in Supplementary Section S1.

The machine-learning energetics plotted in Figure 4 place G1 and G2 below all four perovskite-type lattice structures at every swap fraction. The energetic spread across their enumerated arrangements is notably narrow, spanning only 1.09 and 0.62 meV/atom compared to 4.66 meV/atom for structure B, indicating a dense manifold of near-degenerate states. Lattice structure D instead favors its ordered parent at s = 0, while the cubic lattice structure is strongly destabilized at s = 1 (−3.952 eV/atom), where every Ca is forced onto the corner sublattice. Consistent with the two cation species being physically distinct, the energies are not mirror-symmetric in s even though the arrangement counts obey count(s) = count(1-s).

These separations are smaller than the error of the potential, so the ranking is used for screening only [25,26]. Across the thirty configurations of the six lattice structures carried forward to Table 4, the potential reproduces relative energies with a mean absolute error of about 8 meV/atom and a largest error of 38 meV/atom after aligning the two sets at their means, against lattice-structure separations of a few meV/atom. First-principles relaxation reverses the arrangement ranking within two of the five lattice structures. A machine-learning potential is therefore the right instrument for covering a candidate library exhaustively and the wrong one for deciding between its leading members, a division of labor that transfers to any system whose competing structures lie within the error of the screen.

### 3.2 First-principles energetics, stability against decomposition and hydrogen content

G2 reaches −4.186 eV/atom and G1 −4.184 eV/atom, placing G2 2.10 meV/atom below G1. Table 5 gives those energies for the lowest arrangement of the six lattice structures at every swap fraction, all on one scale. The potential reverses that ordering by 0.44 meV/atom, a difference too fine for it to resolve, so the first-principles ordering stands and G2 is carried forward.

Resolving the arrangement degree of freedom on G2 itself shows its ordered arrangement to be already the ground state of the series. No exchange of Na and Ca lowers the energy, the four rearranged configurations lying +1.08, +1.36, +2.22 and +2.87 meV/atom above it at s = 1, 0.50, 0.25 and 0.75 respectively. Lattice structure B at s = 0.50 lies 7.67 meV/atom higher. Every lattice structure and arrangement examined has a negative formation energy against the elemental standard states (Table 6). G2 is the deepest at −0.3469 eV/atom, deeper than all of the perovskite structures, so the revision of the most stable structure is reproduced in the formation energies as well as in the total energies.

The reaction $NaBH_4 + Ca(BH_4)_2 \rightarrow NaCa(BH_4)_3$ is endothermic for every configuration examined. G2 in its ordered arrangement is the least endothermic at +3.37 meV/atom, and the cubic lattice structure the most at +203.37 meV/atom. The zero-point energy raises the static-lattice value of G2 to +3.64 meV/atom, a shift smaller than a third of a meV per atom. The harmonic free energy of the reaction remains positive up to 1000 K (Figure S5). These same two binary phases are the phases that bound the quaternary convex hull at this composition, so the entries of Table 4 are distances above the hull as well. Figure 5 gives the construction. We conFigure 5structed the quaternary convex hull of Na-Ca-B-H from the 3319 structures of the MPDD that span the system, reduced to the 144 distinct compositions they occupy by retaining the lowest energy at each, all on a single machine-learning energy scale. At $NaCaB_3H_{12}$ the hull is spanned by $NaBH_4$ and $Ca(BH_4)_2$ in the ratio 6:11, so the hull and the binary tie-line coincide exactly rather than merely running parallel (Figure S1). The next-lowest decomposition lies 111.5 meV/atom higher, over twice the largest scatter between energy sets for this system, and the bounding phases are unchanged when the hydrogen-rich phases, least reliably described by the potential, are removed. The machine-learning hull serves only to identify the bounding phases. All Table 4 energies, including those of NaBHTable $4_4$ and $Ca(BH_4)_2$, are DFT values. That endothermicity bounds the release enthalpy of the phase. The +3.37 meV/atom, or +3.64 with zero-point energy, amounts to 5.53 kJ per mole of formula unit over the mixture of the two binaries, 5.97 with zero-point energy. Decomposition to the products those binaries yield requires $NaCa(BH_4)_3 \rightarrow NaH + \frac{1}{2}CaB_6 + \frac{1}{2}CaH_2 + 5H_2$ by element balance, which releases five moles of $H_2$ per formula unit. The enthalpy per mole of $H_2$ then falls 1.1 kJ under the value the same route delivers from the binaries, 1.2 with zero-point energy. That reduction quantifies the destabilization that heterovalent substitution targets. A desorption temperature would need free energies for the three solid products as well, which a separate study takes up.

The same balance fixes what the phase can deliver. Five moles of $H_2$ per formula unit amount to 9.37 wt.% of the 11.24 wt.% the formula holds. G2 holds one formula unit in 153.28 Å$^3$, which puts its crystallographic density at 1.166 g cm$^{-3}$ at 0 K and its full hydrogen content at 131.0 g $H_2$ per liter of crystal. The gravimetric figure equals the one an equimolar mixture of $NaBH_4$ and $Ca(BH_4)_2$ carries, since the ternary holds the same atoms, so composition alone fixes it and no structure can change it.

The energies of Table 5, together with the hull distances of Table 4, describe a dense low-energy landscape rather than one isolated minimum [37]. The five structure types carried to first principles lie between −4.106 and −4.102 eV/atom on the machine-learning surface (Table 3), and relaxation resolves them into three distinct minima between −4.186 and −4.178 eV/atom. G1 and G2 are each reached from two unrelated donors. On G2, every Na/Ca rearrangement lies between +1.08 and +2.87 meV/atom above the ordered arrangement, and the ten G1 and G2 configurations of Table 4 lie between +3.37 and +11.01 meV/atom above the hull. All of these separations lie below $k_BT$ at 300 K, marked in Figure 5(b), This density of near-degenerate minima is why the machine-learning screen cannot fix the ranking, and why each leading candidate had to be relaxed from first principles. Furthermore, the presence of multiple structures within approximately $k_BT$ of one another suggests that $NaCa(BH_4)_3$ may not be adequately described as a single ordered crystal at synthesis temperatures. Instead, configurational contributions may stabilize a manifold of closely related structures. While the present work focuses on ground-state identification, future studies incorporating configurational entropy may be required to determine the finite-temperature equilibrium state.

### 3.3 Dynamical stability

G2 is dynamically stable at every wavevector its supercell resolves exactly and elastically stable in the long-wavelength limit (Figure 6(a)). The $3 \times 2 \times 2$ supercell computes the force constants exactly at twelve wavevectors, and all twelve carry real frequencies only, the lowest away from $\Gamma$ at +1.374 THz (Table S3); the three acoustic branches vanish at $\Gamma$ to within $5 \times 10^{-7}$ THz. The relaxed-ion elastic tensor, obtained by finite strain without a supercell or interpolation, is positive definite, with a lowest eigenvalue of +7.74 GPa (Table S4), so the Born criterion is satisfied as $q \rightarrow 0$. The sound velocities it predicts are consistent with the acoustic slopes of the dispersion (Table S5). Between these two limits, within $|q| \lesssim 0.06$ of $\Gamma$ and inside the first commensurate points at ⅓ a*, ½ b* and ½ c*, the Fourier-interpolated acoustic branch dips to −0.174 THz (−5.8 $cm^{-1}$). This dip is negligible in magnitude, occurs where no frequency is computed directly, and lies between the positive long-wavelength limit and the positive frequencies computed at the commensurate points. A machine-learning test in which only the c repeat of the supercell is doubled removes a comparable near-$\Gamma$ dip, which identifies such features as truncation of the force constants at the supercell boundary (Supplementary Section S2 and Figure S2).

The spectrum of G2 consists of three separated bands (Figure 6): external modes up to 558 $cm^{-1}$, H-B-H bending between 1031 and 1300 $cm^{-1}$ and B-H stretching between 2303 and 2493 $cm^{-1}$.

The lowest optic mode lies at 104.0 cm$^{-1}$. Of the 51 modes per formula unit, 24 are external (three acoustic, twelve translational and nine librational), and 15 and 12 are the internal bending and stretching modes of the three $[BH_4]^-$ units. In the projected density of states (Figure 6(b)), the external modes below 11 THz (367 cm$^{-1}$), 37 % of the total, are shared roughly equally among the cations, boron and hydrogen. Above 11 THz the external modes are librations of the $[BH_4]^-$ units, 94 to 98 % rotational in character at $\Gamma$ and 96 % hydrogen, and the cation contribution falls to about 1 %. The bending and stretching bands are 91 % and 93 % hydrogen and carry no cation contribution.

The harmonic free energy, vibrational entropy and heat capacity of G2 follow from the same force constants [62,64] (Figure 7). The zero-point energy is 315.0 kJ mol$^{-1}$. CombFigure 7ined with the corresponding quantities for $NaBH_4$ and $Ca(BH_4)_2$, they give the reaction free energy of Section 3.2 (Figure S5). These quantities refer to the ordered cation arrangement, whose configurational spread, 2.87 meV/atom (Table 5), is of the same order as the ideal Na/Ca mixing contribution at 300 K, 2.11 meV/atom [65,66].

Neither of the two perovskite-type lattice structures lowest in static energy, B and C, is a local minimum. B at $s = 0.50$ carries one imaginary branch (Figure 8(a)), reaching −0.635 THz (−21.2 cm$^{-1}$) at $q = (½, 0, ½)$, a wavevector commensurate with its 2 × 2 × 2 supercell and therefore computed directly. No other wavevector of an 8 × 8 × 8 mesh carries an imaginary frequency. In the mass-weighted eigenvector, 56 % of the weight is rigid translation of the $[BH_4]^-$ units, 20 % is their libration and 23 % lies on the Ca sublattice, with less than 1 % on Na and on internal deformation. The energy of B decreases along a static distortion that doubles the cell along a and

c. The density of states of B has the same three bands as that of G2 (Figure 8(b)), and its harmonic thermodynamics, with the imaginary branch excluded, are shown in Figure 9.

C at s = 0.50 is unstable along its entire path (Figure S3). All 24 wavevectors of the path are commensurate with its 2 × 2 × 2 supercell, and each carries at least one imaginary mode, reaching −2.401 THz (−80.1 $cm^{-1}$) at q = (½, 0, 0) and −2.351 THz (−78.4 $cm^{-1}$) at Γ. C is therefore a saddle point of the energy surface [67–69]. Of the three lattice structures examined, G2, the lowest in energy, is the only local minimum.

### 3.4 A diffraction fingerprint for G2

A laboratory powder pattern distinguishes G2 from the closest competing lattice structure (Figure 10). Patterns were simulated from the relaxed cells under the instrument model of Section 2.4, with Debye-Waller factors taken from the phonon calculation rather than assumed. They were computed on the same footing for the closest competing lattice structure, because a pattern identifies a phase only in contrast with its closest competitor. Patterns for the three remaining lattice structures are given in Figure S4.

Identification does not rest on the strongest reflection, which for G2 is $1\overline{0}\overline{2}$ at 29.80° (d = 2.996 Å), nor on the number of reflections, although that contrast is large and is the most visible feature of Figure 10. The tick rows give 57 allowed reflections for G2 between 5° and 60° against 170 for B, and the strong reflections of B are 0.65° apart on average against 1.58° for G2. That ratio is fixed by the cell volume divided by the order of the point group, B having twice the volume of G2 and half its symmetry, and would be reproduced by any pair of lattice structures with those metrics.

What identifies G2 is a set of strong reflections beside which the competing lattice structure places no line within twice the instrumental width. Five qualify at $I \geq 10$: 001 at 13.95° (d = 6.345 Å), 010 at 16.52° (5.361 Å), $0\bar{1}\bar{1}$ at 17.69° (5.009 Å), $10\bar{1}$ at 22.07° (4.025 Å) and $0\bar{2}\bar{1}$ at 31.54° (2.834 Å). The 001 and $0\bar{2}\bar{1}$ lines are the most robust, the nearest reflection of B lying 0.64° and 0.56° away. Observing that set, rather than a peak near 29.8°, is what would establish that this lattice structure had formed.

Two limitations are inherent to the technique. X-rays scatter from electrons, so hydrogen, twelve of the seventeen atoms of the G2 cell, contributes almost nothing. The pattern fixes the Na/Ca/B framework but not the orientations of the $[BH_4]^-$ units, for which neutron diffraction on a $^{11}B$-enriched, deuterated sample would be required. Because the cells are relaxed with PBE + D3(BJ), the peak positions carry the systematic error of that functional. A one per cent error in the cell displaces 2θ by roughly 0.3 to 0.5° near 30°, several instrumental widths and larger than the separations quoted here. Those separations are differences between phases computed with the same functional, so the error partly cancels, but it is a further reason to rest an identification on the whole set of lines rather than on any one of them.

### 3.5 Implications

Our results clarify why previous attempts to synthesize equimolar $NaCa(BH_4)_3$ have been unsuccessful. Earlier experimental efforts [27,70] did not yield this composition, prompting suggestions that it lies outside the stability field of complex borohydrides. While our phonon calculations confirm that the previously assumed high-symmetry perovskite configurations are dynamically unstable, the true barrier to synthesis is thermodynamic. As shown in Figure S5, the reaction free energy for $NaBH_4 + Ca(BH4)2 \rightarrow NaCa(BH_4)_3$ remains positive up to 1000 K. In physical synthesis, such as ball-milling, the system naturally relaxes toward the global minimum, which in this case is phase separation into the binary precursors. The failure of earlier syntheses is therefore driven by the thermodynamic metastability of the ternary phase rather than a limitation of specific structure types. Should non-equilibrium synthesis techniques succeed in overcoming this barrier, the characteristic X-ray diffraction reflections identified in Section 3.4 will provide the direct experimental fingerprint needed to confirm the G2 phase.

A synthetic route follows from it. The perovskite-type $MSr(BH_4)_3$ borohydrides (M = K, Rb, Cs) were obtained by mechanochemical addition of $MBH_4$ to $Sr(BH_4)_2$ [28], and the corresponding addition of $NaBH_4$ to $Ca(BH_4)_2$ is the natural first attempt at this composition. Annealing under a hydrogen overpressure and crystallization from a coordinating solvent followed by desolvation are the established alternatives within this family [71]. Whether such an attempt succeeds is a kinetic question rather than a thermodynamic one, because at +3.37 meV/atom above the tie-line the lattice structure is metastable with respect to its own parents. The same 3.37 meV/atom lowers its release enthalpy by 1.1 kJ per mole of $H_2$, so the phase that resists synthesis is also the one that would release hydrogen more readily than its precursors. This metastability does not preclude synthesis [72,73]. The median metastability of 29 902 observed inorganic crystalline phases is 15 ± 0.5 meV/atom [74], over four times the value here. The $Na_{1-x}K_xBH_4$ solid solution, also above its parents, forms above 110 °C and persists for hours at room temperature before separating into NaBH4 and $KBH_4$ [75]. $NaCa(BH_4)_3$ is thus best targeted by high-temperature synthesis, rapid cooling and prompt characterization.

# 4 Conclusion

$NaCa(BH_4)_3$ has no reported crystal structure, and every candidate previously considered for it was a perovskite. Its most stable form is not. An unrestricted search over 2238 candidates and 118 inequivalent cation arrangements places a monoclinic Cm structure 7.67 meV/atom below the best perovskite-type candidate. Neither template nor arrangement decides it. Two unrelated donors agree to within 0.076 meV/atom, and its ordered arrangement is the ground state of a 2.87 meV/atom series. Against the two binary borohydrides, which bound the quaternary hull, it is metastable by +3.37 meV/atom, so thermodynamics does not forbid it. The same margin places its release enthalpy 1.1 kJ per mole of $H_2$ under that of the two binaries along the same decomposition route.

G2 carries no imaginary mode at any wavevector its supercell resolves exactly, the lowest at +1.374 THz, and its relaxed-ion elastic tensor is positive definite. The two nearest perovskite-type candidates fail. B retains a zone-boundary mode on the $[BH_4]^-$ sublattice, and C is imaginary at every wavevector of its path, reaching −2.401 THz. Lattice dynamics thus decide what the static energies cannot. **However, while G2 is dynamically stable, its thermodynamic metastability against phase separation (+3.37 meV/atom) explains why standard equilibrium syntheses have not produced it.**

Across the thirty configurations of the six lattice structures carried forward to Table 4, the potential reproduces them to 8 meV/atom. That error is larger than the few meV/atom separating the leading candidates and enough to reverse C1 and C2, so that G2, not G1, is the ground state at this composition. The potential covers a library exhaustively but cannot choose among its leaders.

What is established here is structural and static. Hydrogen release temperatures and the equilibrium decomposition path lie outside it. They require the same vibrational treatment for every species in a candidate reaction, with the cation sublattice as a statistical ensemble, whose 2.87 meV/atom spread matches the 2.11 meV/atom of ideal mixing at 300 K. More generally, this work establishes a workflow for unknown crystal structures in complex hydrides. A database-wide search and exhaustive cation enumeration feed first-principles validation and lattice-dynamics verification, and the outcome shows that a search confined to perovskite-type candidates would have missed the ground state.

The low-energy landscape of this composition is dense, with competing minima separated by only a few meV/atom, so finite-temperature effects may decide which of them forms. The simulated diffraction patterns give specific targets for experimental identification and synthesis. Five strong reflections of G2 have no line of B within twice the instrumental width, and the 001 line at 13.95° and the $0\overline{21}$ line at 31.54° are the most robust. Observing them in the products of the mechanochemical or high-temperature routes of Section 3.5 would show that this phase has formed.

**CRediT authorship contribution statement**

**Sanaa Ismail:** Conceptualization, Methodology, Software, Formal analysis, Investigation, Data curation, Visualization, Writing (original draft). **Ricardo Amaral:** Conceptualization, Visualization, Software, Writing (review and editing). **Attia A. Gadallah:** Software, Investigation, Methodology, Visualization, Writing (review and editing). **Hassan Mohamed El-Said Azzazy:** Supervision, Writing (review and editing). **Zi-Kui Liu:** Conceptualization, Supervision, Resources, Writing (review and editing).

**Declaration of competing interest**

The authors declare that they have no known competing financial interests or personal relationships that could have appeared to influence the work reported in this paper.

**Funding**

This research did not receive any specific grant from funding agencies in the public, commercial, or not-for-profit sectors. This work used Bridges-2 at the Pittsburgh Supercomputing Center through allocation DMR190011 from the Advanced Cyberinfrastructure Coordination Ecosystem: Services & Support (ACCESS) program, which is supported by U.S. National Science Foundation grants #2138259, #2138286, #2138307, #2137603, and #2138296

**Data availability**

The first-principles crystal structure of G2 is openly available in Zenodo at https://doi.org/10.5281/zenodo.23377002, as CIF files of the conventional (Cm) and primitive cells together with the corresponding VASP structure file and a README file describing the calculation settings.

## References

[1] Hannan MA, Hoque MM, Mohamed A, Ayob A. Review of energy storage systems for electric vehicle applications: Issues and challenges. Renewable and Sustainable Energy Reviews 2017;69:771–89. https://doi.org/10.1016/j.rser.2016.11.171.

[2] Kittner N, Lill F, Kammen DM. Energy storage deployment and innovation for the clean energy transition. Nat Energy 2017;2:17125. https://doi.org/10.1038/nenergy.2017.125.

[3] Ould Amrouche S, Rekioua D, Rekioua T, Bacha S. Overview of energy storage in renewable energy systems. Int J Hydrogen Energy 2016;41:20914–27. https://doi.org/10.1016/j.ijhydene.2016.06.243.

[4] Ismail S, Khedr GE, Salem FZ, Allam NK. Defect-Engineered SrZrO3: Unraveling the Role of Nitrogen and Carbon in Optoelectronic and Catalytic Performance. Energy and Fuels 2025;39:13703–14. https://doi.org/10.1021/acs.energyfuels.5c01985.

[5] Wolf E. Large-Scale Hydrogen Energy Storage. Electrochemical Energy Storage for Renewable Sources and Grid Balancing 2015:129–42. https://doi.org/10.1016/B978-0-444-62616-5.00009-7.

[6] Allendorf MD, Stavila V, Snider JL, Witman M, Bowden ME, Brooks K, et al. Challenges to developing materials for the transport and storage of hydrogen. Nat Chem 2022;14:1214–23. https://doi.org/10.1038/s41557-022-01056-2.

[7] Bellon Monsalve D, Kefi C, Klopcic N, Huot J. Review of hydrogen storage using AB, AB2, HEA, and solid solution metal hydrides. Int J Hydrogen Energy 2026;229:154686. https://doi.org/10.1016/j.ijhydene.2026.154686.

[8] Arsad AZ, Hannan MA, Al-Shetwi AQ, Mansur M, Muttaqi KM, Dong ZY, et al. Hydrogen energy storage integrated hybrid renewable energy systems: A review analysis for future research directions. Int J Hydrogen Energy 2022;47:17285–312. https://doi.org/10.1016/j.ijhydene.2022.03.208.

[9] Xu Y, Zhou Y, Li Y, Ding Z. Research Progress and Application Prospects of Solid-State Hydrogen Storage Technology. Molecules 2024;29:1767. https://doi.org/10.3390/molecules29081767.

[10] Bösenberg U, Kim JW, Gosslar D, Eigen N, Jensen TR, von Colbe JMB, et al. Role of additives in LiBH4–MgH2 reactive hydride composites for sorption kinetics. Acta Mater 2010;58:3381–9. https://doi.org/10.1016/j.actamat.2010.02.012.

[11] Hummelshøj JS, Landis DD, Voss J, Jiang T, Tekin A, Bork N, et al. Density functional theory based screening of ternary alkali-transition metal borohydrides: A computational material design project. J Chem Phys 2009;131:014101. https://doi.org/10.1063/1.3148892.

[12] Xiong Z, Yong CK, Wu G, Chen P, Shaw W, Karkamkar A, et al. High-capacity hydrogen storage in lithium and sodium amidoboranes. Nat Mater 2008;7:138–41. https://doi.org/10.1038/nmat2081.

[13] Abdechafik E harrak, Ait Ousaleh H, Mehmood S, Filali Baba Y, Bürger I, Linder M, et al. An analytical review of recent advancements on solid-state hydrogen storage. Int J Hydrogen Energy 2024;52:1182–93. https://doi.org/10.1016/j.ijhydene.2023.10.218.

[14] Nakamori Y, Miwa K, Ninomiya A, Li H, Ohba N, Towata S, et al. Correlation between thermodynamical stabilities of metal borohydrides and cation electronegativites: First-principles calculations and experiments. Phys Rev B 2006;74:045126. https://doi.org/10.1103/PhysRevB.74.045126.

[15] Ozolins V, Majzoub EH, Wolverton C. First-Principles Prediction of Thermodynamically Reversible Hydrogen Storage Reactions in the Li-Mg-Ca-B-H System. J Am Chem Soc 2009;131:230–7. https://doi.org/10.1021/ja8066429.

[16] Guo Y, Ren Y, Wu H, Jia J. Prediction of thermodynamically reversible hydrogen storage reactions utilizing Ca–M(M = Li, Na, K)–B–H systems: a first-principles study. J Mol Model 2013;19:5135–42. https://doi.org/10.1007/s00894-013-2012-8.

[17] Zhou P, Zhou Q, Xiao X, Fan X, Zou Y, Sun L, et al. Machine Learning in Solid-State Hydrogen Storage Materials: Challenges and Perspectives. Adv Mater 2025;37:2413430. https://doi.org/10.1002/adma.202413430.

[18] Krajewski AM, Siegel JW, Xu J, Liu Z-K. Extensible Structure-Informed Prediction of Formation Energy with improved accuracy and usability employing neural networks. Comput Mater Sci 2022;208:111254. https://doi.org/10.1016/j.commatsci.2022.111254.

[19] Ward L, Liu R, Krishna A, Hegde VI, Agrawal A, Choudhary A, et al. Including crystal structure attributes in machine learning models of formation energies via Voronoi tessellations. Phys Rev B 2017;96:024104. https://doi.org/10.1103/PhysRevB.96.024104.

[20] Zhou Q, Jiang M, Xu J, Xie Z-H, Munroe P. Accelerated discovery of hydrogen storage hydride perovskites: A combined machine learning and first-principles approach. Int J Hydrogen Energy 2025;179:151697. https://doi.org/10.1016/j.ijhydene.2025.151697.

[21] Wolverton C. Crystal structure and stability of complex precipitate phases in Al–Cu–Mg–(Si) and Al–Zn–Mg alloys. Acta Mater 2001;49:3129–42. https://doi.org/10.1016/S1359-6454(01)00229-4.

[22] Ravi C, Wolverton C. First-principles study of crystal structure and stability of Al-Mg-Si-(Cu) precipitates. Acta Mater 2004;52:4213–27. https://doi.org/10.1016/j.actamat.2004.05.037.

[23] Zhong Y, Wolverton C, Austin Chang Y, Liu Z-K. A combined CALPHAD/first-principles remodeling of the thermodynamics of Al–Sr: unsuspected ground state energies by "rounding up the (un)usual suspects." Acta Mater 2004;52:2739–54. https://doi.org/10.1016/j.actamat.2004.02.022.

[24] Mishin Y. Machine-learning interatomic potentials for materials science. Acta Mater 2021;214:116980. https://doi.org/10.1016/j.actamat.2021.116980.

[25] Zhu S, Sarıtürk D, Arróyave R. Accelerating CALPHAD-based phase diagram predictions in complex alloys using universal machine learning potentials: Opportunities and challenges. Acta Mater 2025;286:120747. https://doi.org/10.1016/j.actamat.2025.120747.

[26] Liu Y, Mo Y. Assessing the accuracy of machine learning interatomic potentials in predicting the elemental orderings: A case study of Li-Al alloys. Acta Mater 2024;268:119742. https://doi.org/10.1016/j.actamat.2024.119742.

[27] Schouwink P, Morelle F, Sadikin Y, Filinchuk Y, Černý R. Increasing hydrogen density with the cation-anion pair BH4--NH4+ in Perovskite-Type NH4Ca(BH4)3. Energies 2015;8:8286–99. https://doi.org/10.3390/en8088286.

[28] Møller KT, Ley MB, Schouwink P, Černý R, Jensen TR. Synthesis and thermal stability of perovskite alkali metal strontium borohydrides. Dalton Transactions 2016;45:831–40. https://doi.org/10.1039/C5DT03590B.

[29] Merchant A, Batzner S, Schoenholz SS, Aykol M, Cheon G, Cubuk ED. Scaling deep learning for materials discovery. Nature 2023;624:80–5. https://doi.org/10.1038/s41586-023-06735-9.

[30] Ong SP, Richards WD, Jain A, Hautier G, Kocher M, Cholia S, et al. Python Materials Genomics (pymatgen): A robust, open-source python library for materials analysis. Comput Mater Sci 2013;68:314–9. https://doi.org/10.1016/j.commatsci.2012.10.028.

[31] Buerger MJ. Derivative Crystal Structures. J Chem Phys 1947;15:1–16. https://doi.org/10.1063/1.1746278.

[32] Hart GLW, Forcade RW. Algorithm for generating derivative structures. Phys Rev B 2008;77:224115. https://doi.org/10.1103/PhysRevB.77.224115.

[33] Hart GLW, Forcade RW. Generating derivative structures from multilattices: Algorithm and application to hcp alloys. Phys Rev B 2009;80:014120. https://doi.org/10.1103/PhysRevB.80.014120.

[34] Ångqvist M, Muñoz WA, Rahm JM, Fransson E, Durniak C, Rozyczko P, et al. icet - A Python library for constructing and sampling alloy cluster expansions. Adv Theory Simul 2019;2:1900015. https://doi.org/10.1002/adts.201900015.

[35] Hart GLW, Nelson LJ, Forcade RW. Generating derivative structures at a fixed concentration. Comput Mater Sci 2012;59:101–7. https://doi.org/10.1016/j.commatsci.2012.02.015.

[36] Bragg WL, Williams EJ. The effect of thermal agitation on atomic arrangement in alloys. Proceedings of the Royal Society of London Series A, Containing Papers of a Mathematical and Physical Character 1934;145:699–730. https://doi.org/10.1098/rspa.1934.0132.

[37] Oganov AR, Valle M. How to quantify energy landscapes of solids. J Chem Phys 2009;130:104504. https://doi.org/10.1063/1.3079326.

[38] Valle M, Oganov AR. Crystal fingerprint space - A novel paradigm for studying crystal-structure sets. Acta Crystallogr A 2010;66:507–17. https://doi.org/10.1107/S0108767310026395.

[39] Van de Walle A, Asta M, Ceder G. The alloy theoretic automated toolkit: A user guide. Calphad 2002;26:539–53. https://doi.org/10.1016/S0364-5916(02)80006-2.

[40] van de Walle A, Ceder G. Automating first-principles phase diagram calculations. J Phase Equilib 2002;23:348–59. https://doi.org/10.1361/105497102770331596.

[41] van de Walle A. Multicomponent multisublattice alloys, nonconfigurational entropy and other additions to the Alloy Theoretic Automated Toolkit. Calphad 2009;33:266–78. https://doi.org/10.1016/j.calphad.2008.12.005.

[42] Batatia I, Kovács DP, Simm GNC, Ortner C, Csányi G. MACE: Higher Order Equivariant Message Passing Neural Networks for Fast and Accurate Force Fields. Adv Neural Inf Process Syst 2022;35:11423–36. https://doi.org/10.48550/arXiv.2206.07697.

[43] Batatia I, Benner P, Chiang Y, Elena AM, Kovács DP, Riebesell J, et al. A foundation model for atomistic materials chemistry. J Chem Phys 2025;163. https://doi.org/10.1063/5.0297006.

[44] Hjorth Larsen A, Jørgen Mortensen J, Blomqvist J, Castelli IE, Christensen R, Dułak M, et al. The atomic simulation environment—a Python library for working with atoms. Journal of Physics: Condensed Matter 2017;29:273002. https://doi.org/10.1088/1361-648X/aa680e.

[45] Kresse G, Hafner J. *Ab initio* molecular dynamics for liquid metals. Phys Rev B 1993;47:558–61. https://doi.org/10.1103/PhysRevB.47.558.

[46] Kresse G, Furthmüller J. Efficiency of ab-initio total energy calculations for metals and semiconductors using a plane-wave basis set. Comput Mater Sci 1996;6:15–50. https://doi.org/10.1016/0927-0256(96)00008-0.

[47] Kresse G, Furthmüller J. Efficient iterative schemes for *ab initio* total-energy calculations using a plane-wave basis set. Phys Rev B 1996;54:11169–86. https://doi.org/10.1103/PhysRevB.54.11169.

[48] Blöchl PE. Projector augmented-wave method. Phys Rev B 1994;50:17953–79. https://doi.org/10.1103/PhysRevB.50.17953.

[49] Kresse G, Joubert D. From ultrasoft pseudopotentials to the projector augmented-wave method. Phys Rev B 1999;59:1758–75. https://doi.org/10.1103/PhysRevB.59.1758.

[50] Perdew JP, Burke K, Ernzerhof M. Generalized Gradient Approximation Made Simple. Phys Rev Lett 1996;77:3865–8. https://doi.org/10.1103/PhysRevLett.77.3865.

[51] Grimme S, Ehrlich S, Goerigk L. Effect of the damping function in dispersion corrected density functional theory. J Comput Chem 2011;32:1456–65. https://doi.org/10.1002/jcc.21759.

[52] Grimme S, Antony J, Ehrlich S, Krieg H. A consistent and accurate *ab initio* parametrization of density functional dispersion correction (DFT-D) for the 94 elements H-Pu. J Chem Phys 2010;132:154104. https://doi.org/10.1063/1.3382344.

[53] Monkhorst HJ, Pack JD. Special points for Brillouin-zone integrations. Phys Rev B 1976;13:5188–92. https://doi.org/10.1103/PhysRevB.13.5188.

[54] Jain A, Ong SP, Hautier G, Chen W, Richards WD, Dacek S, et al. Commentary: The materials project: A materials genome approach to accelerating materials innovation. APL Mater 2013;1:011002. https://doi.org/10.1063/1.4812323.

[55] Togo A, Shinohara K, Tanaka I. Spglib: a software library for crystal symmetry search. Sci Technol Adv Mater Methods 2024;4:2384822. https://doi.org/10.1080/27660400.2024.2384822.

[56] Togo A, Tanaka I. First principles phonon calculations in materials science. Scr Mater 2015;108:1–5. https://doi.org/10.1016/j.scriptamat.2015.07.021.

[57] Togo A. First-principles Phonon Calculations with Phonopy and Phono3py. J Phys Soc Jpn 2023;92:012001. https://doi.org/10.7566/JPSJ.92.012001.

[58] Gadallah AA. hydrophonokit: a material-aware scientific framework for automated VASP/Phonopy phonon calculations (version 2.7.0). Zenodo; 2026. https://doi.org/10.5281/zenodo.20337121.

[59] Gajdoš M, Hummer K, Kresse G, Furthmüller J, Bechstedt F. Linear optical properties in the projector-augmented wave methodology. Phys Rev B 2006;73:045112. https://doi.org/10.1103/PhysRevB.73.045112.

[60] Gonze X, Lee C. Dynamical matrices, Born effective charges, dielectric permittivity tensors, and interatomic force constants from density-functional perturbation theory. Phys Rev B 1997;55:10355–68. https://doi.org/10.1103/PhysRevB.55.10355.

[61] Togo A, Chaput L, Tadano T, Tanaka I. Implementation strategies in phonopy and phono3py. J Phys Condens Matter 2023;35:353001. https://doi.org/10.1088/1361-648X/acd831.

[62] Shang S-L, Hector LG, Shi S, Qi Y, Wang Y, Liu Z-K. Lattice dynamics, thermodynamics and elastic properties of monoclinic Li2CO3 from density functional theory. Acta Mater 2012;60:5204–16. https://doi.org/10.1016/j.actamat.2012.06.006.

[63] Verdal N, Hartman MR, Jenkins T, DeVries DJ, Rush JJ, Udovic TJ. Reorientational dynamics of NaBH4 and KBH4. J Phys Chem C 2010;114:10027–33. https://doi.org/10.1021/jp1006473.

[64] Wang Y, Liu Z-K, Chen L-Q. Thermodynamic properties of Al, Ni, NiAl, and Ni3Al from first-principles calculations. Acta Mater 2004;52:2665–71. https://doi.org/10.1016/j.actamat.2004.02.014.

[65] Liu Z-K. Computational thermodynamics and its applications. Acta Mater 2020;200:745–92. https://doi.org/10.1016/j.actamat.2020.08.008.

[66] Lumley SC, Grimes RW, Murphy ST, Burr PA, Chroneos A, Chard-Tuckey PR, et al. The thermodynamics of hydride precipitation: The importance of entropy, enthalpy and disorder. Acta Mater 2014;79:351–62. https://doi.org/10.1016/j.actamat.2014.07.019.

[67] Patrick CE, Jacobsen KW, Thygesen KS. Anharmonic stabilization and band gap renormalization in the perovskite CsSnI3. Phys Rev B 2015;92:201205. https://doi.org/10.1103/PhysRevB.92.201205.

[68] Pallikara I, Kayastha P, Skelton JM, Whalley LD. The physical significance of imaginary phonon modes in crystals. Electron Struct 2022;4:033002. https://doi.org/10.1088/2516-1075/ac78b3.

[69] Zhang H, Dai J, Cao Y, Zhang Y, Bao M, Yin Y. Influence of Alloying Elements on the Phase Structure, Stress–Strain Behavior, and Fracture Toughness of Ni3Sn: A First-Principles Study. Materials 2025;18:1792. https://doi.org/10.3390/ma18081792.

[70] Schouwink P, Ley MB, Tissot A, Hagemann H, Jensen TR, Smrčok Ľ, et al. Structure and properties of complex hydride perovskite materials. Nat Commun 2014;5:5706. https://doi.org/10.1038/ncomms6706.

[71] Paskevicius M, Jepsen LH, Schouwink P, Černý R, Ravnsbæk DB, Filinchuk Y, et al. Metal borohydrides and derivatives – synthesis, structure and properties. Chem Soc Rev 2017;46:1565–634. https://doi.org/10.1039/C6CS00705H.

[72] Edalati K, Emami H, Staykov A, Smith DJ, Akiba E, Horita Z. Formation of metastable phases in magnesium–titanium system by high-pressure torsion and their hydrogen storage performance. Acta Mater 2015;99:150–6. https://doi.org/10.1016/j.actamat.2015.07.060.

[73]Amaral R, Liu N, Poudel B, Li W, Liu Z-K. Ab-initio CALPHAD modeling of NaNH2BH3 for solid-state H2 storage applications. Int J Hydrogen Energy 2025;166:150807. https://doi.org/10.1016/j.ijhydene.2025.150807.

[74]Sun W, Dacek ST, Ong SP, Hautier G, Jain A, Richards WD, et al. The thermodynamic scale of inorganic crystalline metastability. Sci Adv 2016;2:e1600225. https://doi.org/10.1126/sciadv.1600225.

[75]Jensen SRH, Jepsen LH, Skibsted J, Jensen TR. Phase Diagram for the $NaBH_4$–$KBH_4$ System and the Stability of a $Na_{1-x}K_xBH_4$ Solid Solution. J Phys Chem C 2015;119:27919–29. https://doi.org/10.1021/acs.jpcc.5b09851.

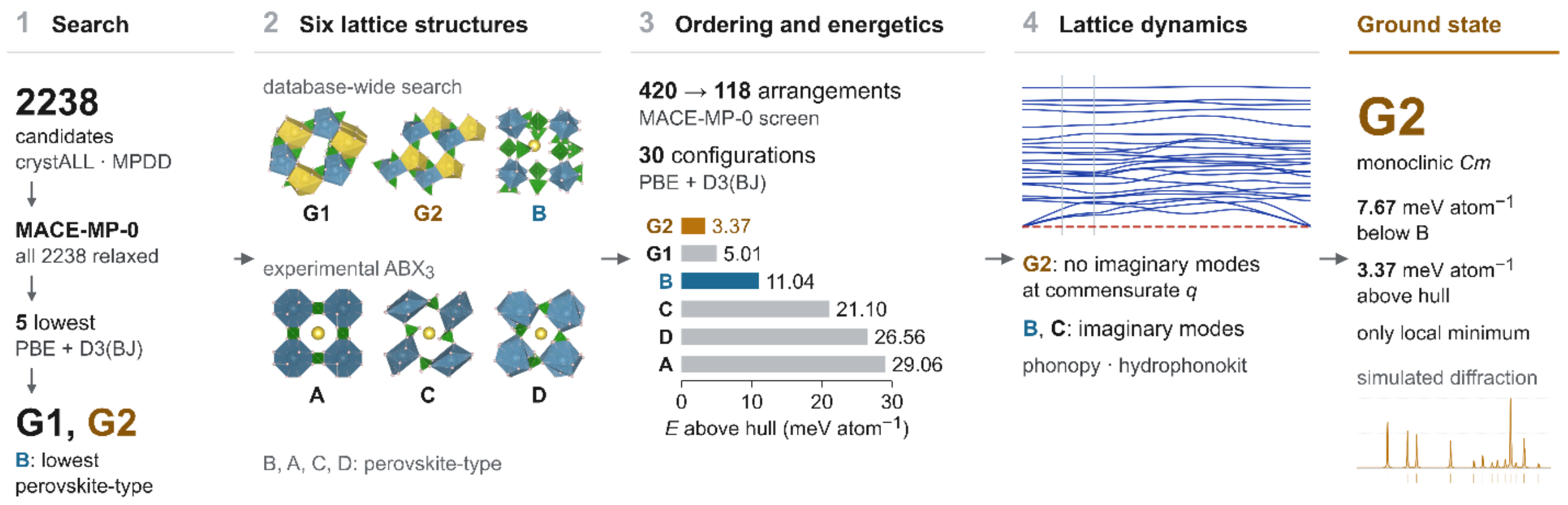

***Figure 1. Workflow used to identify the crystal structure of $NaCa(BH_4)_3$. (1) Search. crystALL retrieves 2238 candidates from the MPDD, MACE-MP-0 relaxes all of them, and the five lowest are relaxed with PBE + D3(BJ). They yield G1 and G2, and B is the lowest-energy perovskite-type candidate of the search. (2) The six lattice structures carried forward, G1, G2 and B from the database-wide search and A, C and D from experimentally determined $ABX_3$ borohydrides. (3) Ordering and energetics. The 420 Na/Ca arrangements reduce to 118 symmetry-inequivalent ones for the MACE-MP-0 screen, and 30 configurations are relaxed with PBE + D3(BJ). Bars give the distance above the convex hull of the lowest arrangement of each lattice structure (Table 4). (4) Lattice dynamics with phonopy and hydrophonokit. G2 has no imaginary mode at the wavevectors its supercell resolves exactly, whereas B and C carry imaginary modes. The last panel summarizes the ground state, G2 (monoclinic Cm), which lies 7.67 meV/atom below B and 3.37 meV/atom above the hull and is the only local minimum of the three lattice structures whose phonons were computed, with its simulated powder diffraction pattern.***

***Table 1. Provenance and relaxed structure of the candidate lattice structures. Space groups (symmetry tolerance 0.05 Å) and atomic volumes refer to the first-principles relaxed structures (Section 2.3) in their ordered arrangement (s = 0), and cell sizes to the native cell of each lattice structure.***

| **Candidate ID** | **Structure Type** | **Cell (atoms)** | **Space group** | **V/atom (Å³)** | **Donors References** |
|---|---|---|---|---|---|
| A | Cubic | 17 | $P\bar{4}3m$ | 11.500 | $NH_4Ca(BH_4)_3$ [27] |
| C | Orthorhombic | 34 | $P2_12_12$ | 8.896 | $CsSr(BH_4)_3$ [28] |
| D | Orthorhombic | 68 | $Pna2_1$ | 10.099 | $RbSr(BH_4)_3$ [28] |
| B | Distorted perovskite | 34 | $P\bar{1}$ | 8.857 | $NbV(PO_4)_3$ (GNoME a9b189292f) [29] |
| G1 | Orthorhombic | 17 | P1 | 9.028 | $Lu_{12}ZnFe_3Sb$ (GNoME 28953a3c7c) [29] |
| G2 | Monoclinic | 17 | Cm | 9.016 | $Er_{12}Fe_3AgSn$ (GNoME 8aecf8173f) [29] |

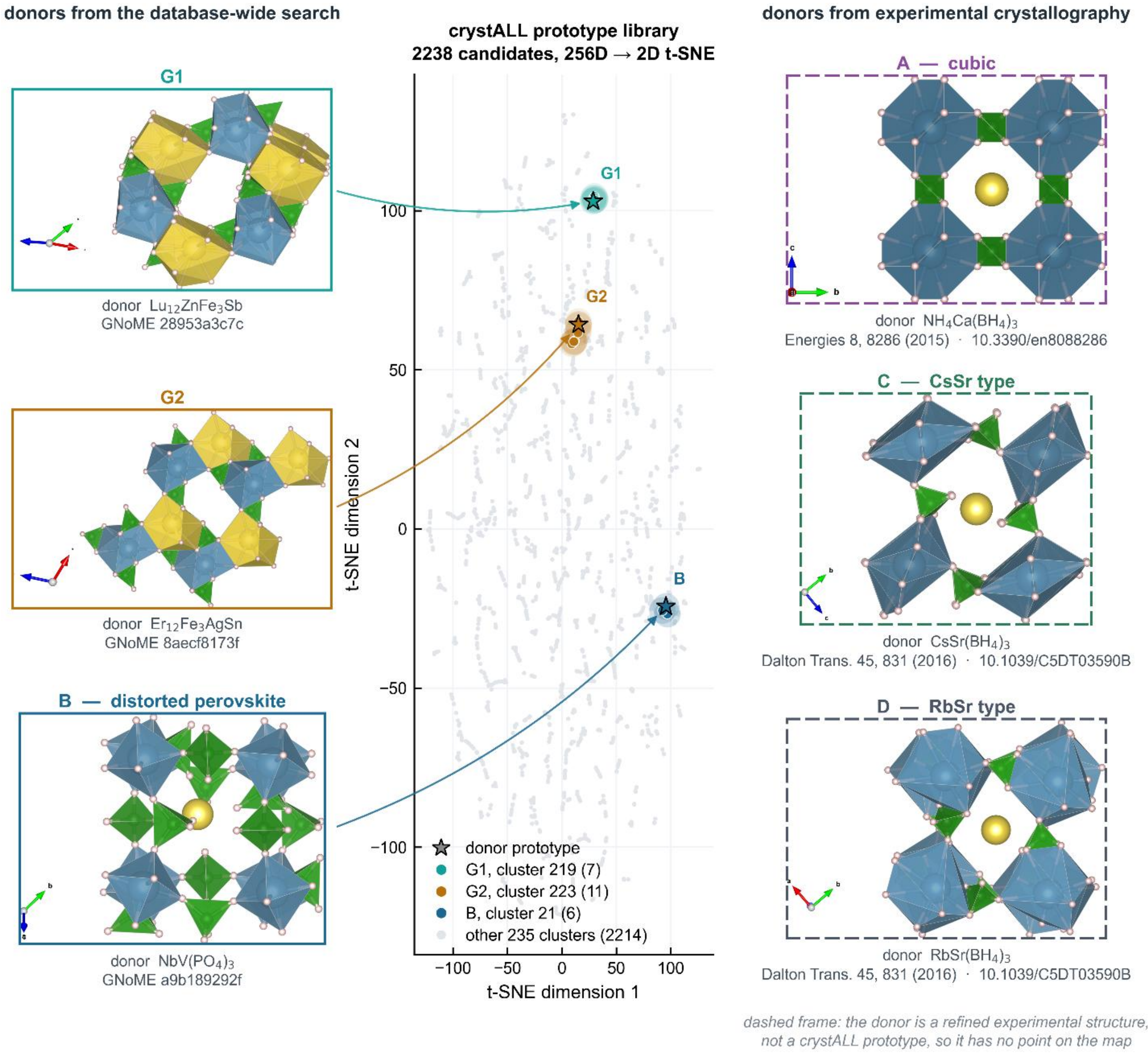

***Figure 2. Provenance of the six candidate lattice structures. Left: G1, G2 and B, recovered by the database-wide search, each with its crystALL donor prototype. Arrows locate each donor on the map of Figure 3, with the donor cluster colored around it. Right: A, C and D, taken from experimentally determined $ABX_3$ borohydrides. Their donors are refined experimental structures, not crystALL prototypes, so they have no map position or arrow (dashed frames). Ca-centered coordination polyhedra are blue and $[BH_4]^-$ tetrahedra green throughout. Na is a yellow sphere, except in G1 and G2, where it is a yellow polyhedron.***

***Table 2. Symmetry-inequivalent Na/Ca arrangements (raw → unique) at each swap fraction s, at fixed equimolar composition. The raw counts are identical for every lattice structure, so the unique counts reflect lattice-structure symmetry alone. All cells carry eight cation sites: B and C as 1×1×2 supercells, A, G1, and G2 as 1×2×2, D natively.***

| **Candidate ID** | **s = 0** | **s = 0.25** | **s = 0.50** | **s = 0.75** | **s = 1** | **Raw → Unique** |
|---|---|---|---|---|---|---|
| A | 1 → 1 | 16 → 1 | 36 → 1 | 16 → 1 | 1 → 1 | 70 → 5 |
| B | 1 → 1 | 16 → 8 | 36 → 20 | 16 → 8 | 1 → 1 | 70 → 38 |
| C | 1 → 1 | 16 → 3 | 36 → 7 | 16 → 3 | 1 → 1 | 70 → 15 |
| D | 1 → 1 | 16 → 4 | 36 → 12 | 16 → 4 | 1 → 1 | 70 → 22 |
| G1 | 1 → 1 | 16 → 4 | 36 → 9 | 16 → 4 | 1 → 1 | 70 → 19 |
| G2 | 1 → 1 | 16 → 4 | 36 → 9 | 16 → 4 | 1 → 1 | 70 → 19 |
| **Total** | **6 → 6** | **96 → 24** | **216 → 58** | **96 → 24** | **6 → 6** | **420 → 118** |

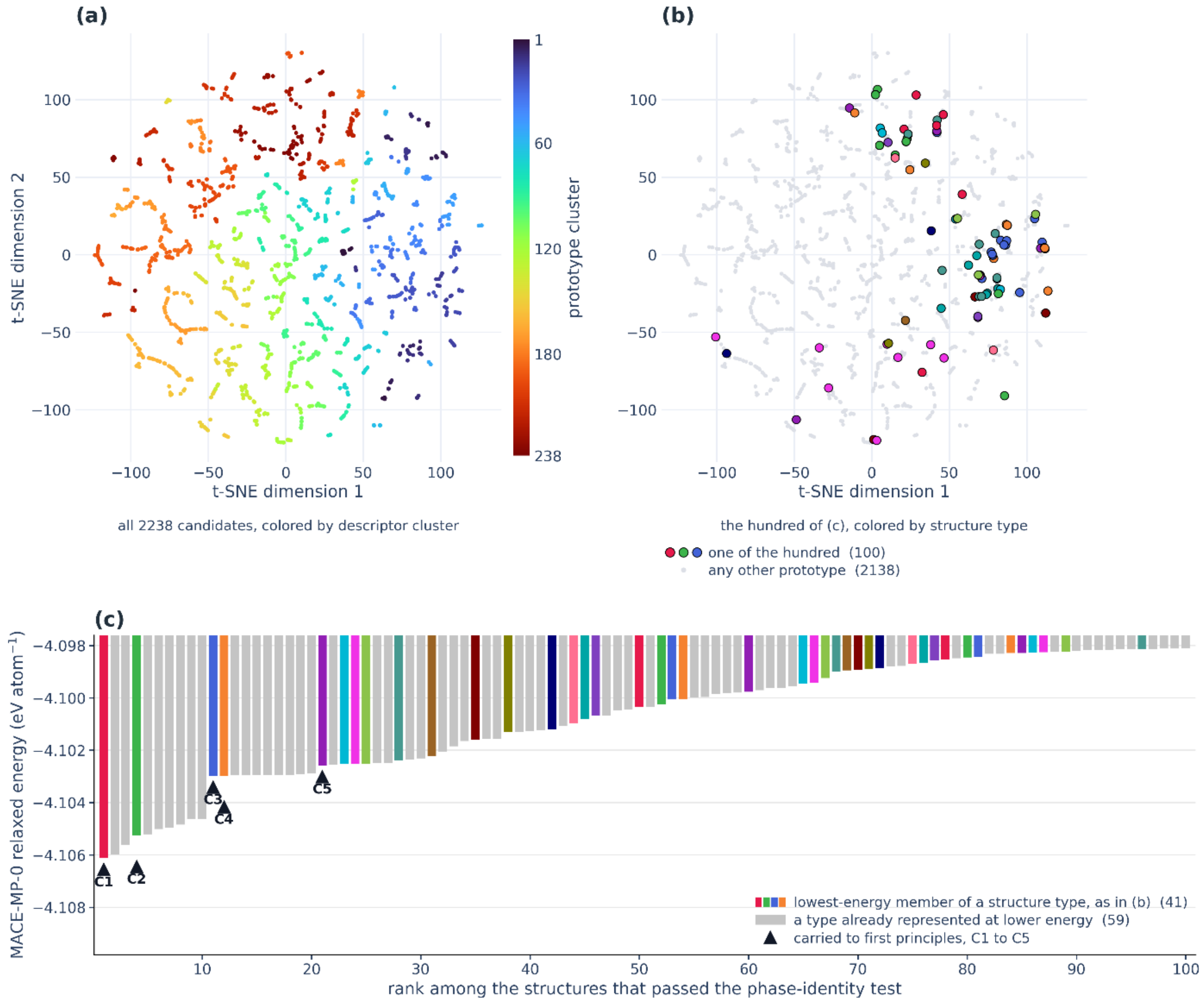

*Figure 3. Outcome of the database-wide search on the crystALL prototype map. (a) All 2238 candidates on the two-dimensional t-SNE embedding of their descriptors, colored by prototype cluster. The color scale at the right covers the 238 clusters and applies to this panel only. (b) The hundred structures of (c) on the same embedding, colored by structure type assigned after relaxation, with all other prototypes in pale gray. (c) Those hundred structures ranked by MACE-MP-0 relaxed energy. A bar takes its color from (b) when it is the lowest-energy member of its structure type and is gray when that type already appears at lower energy. The five structure types carried to first principles, C1-C5, are marked. Panel (a) shows descriptor*

***clusters computed before relaxation and panel (b) structure types assigned after it, so the two color scales are not comparable.***

*Table 3. The five candidates carried to first-principles relaxation.*

| **Donors** | **$Lu_{12}ZnFe_3Sb$** | **$Er_{12}Fe_3AgSn$** | **$SmGd(SeO_4)_3$** | **$Tm_{12}TlFe_3Bi$** | **$Lu_{12}InFe_3Bi$** |
|---|---|---|---|---|---|
| ***CrystALL*** Candidate | C1 | C2 | C3 | C4 | C5 |
| Resolves to (after DFT) | G1 | G2 | G3 | G2 | G1 |
| Database identifier | GNoME 28953a3c7c | GNoME 8aecf8173f | GNoME f5ac6f46a5 | GNoME bf67ce5d08 | GNoME 56655fb175 |
| Permutation | p1 | p1 | p2 | p2 | p1 |
| Donor crystal system | orthorhombic | orthorhombic | trigonal | orthorhombic | orthorhombic |
| Atoms per cell | 17 | 17 | 34 | 17 | 17 |
| MACE-MP-0 energy (eV/atom) | −4.106 | −4.105 | −4.103 | −4.103 | −4.102 |
| MACE-MP-0 volume (Å³/atom) | 9.902 | 9.869 | 9.327 | 9.791 | 9.790 |
| MACE-MP-0 space group | P1 | P1 | $R\bar{3}$ | P1 | P1 |
| Median B-H (Å) | 1.229 | 1.229 | 1.225 | 1.230 | 1.228 |
| DFT energy (eV/atom) | −4.184 | −4.186 | −4.178 | −4.185 | −4.184 |

C1 to C5 are the lowest-energy members of the five lowest structure types of the machine-learning screen.

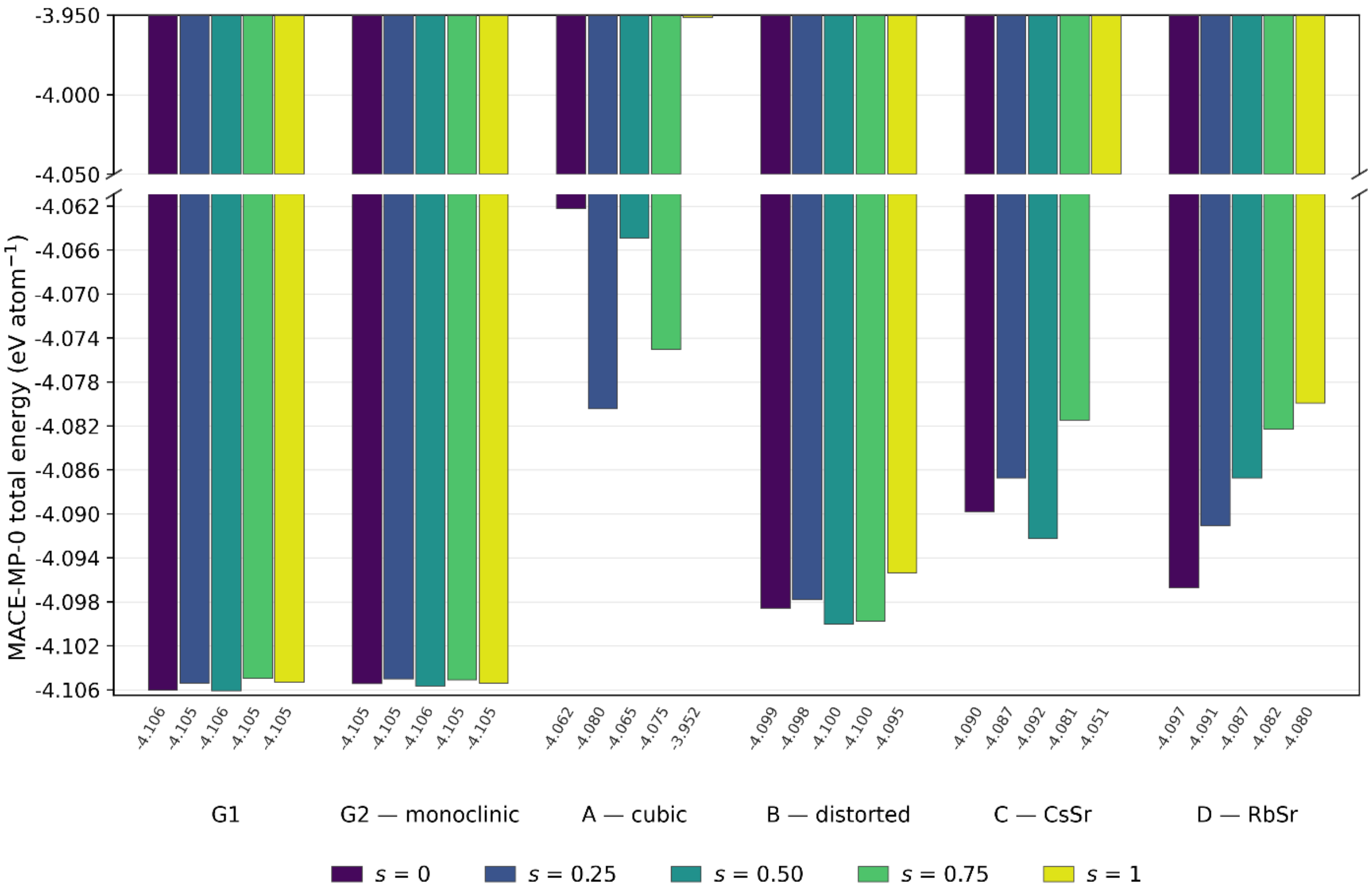

***Figure 4. Machine-learning energetics of the cation-ordering enumeration. Lowest MACE-MP-0 relaxed energy of each lattice structure at each swap fraction s, plotted as total energy so that no configuration serves as zero. Color runs with s from the ordered parent (s = 0) to its complete Na↔Ca complement (s = 1), and the two lattice structures from the database-wide search come first. G1 and G2 lie below all four perovskite-type lattice structures at every swap fraction and span only 1.09 and 0.62 meV/atom, against 4.66 meV/atom for B. Cubic A and CsSr-type C at s = 1 lie far above the common range, at −3.952 and −4.051 eV atom⁻¹, in the compressed upper panel of the broken axis, labeled with their true values.***

***Table 4. Reaction energy of $NaCa(BH_4)_3$ with respect to $NaBH_4$ + $Ca(BH_4)_2$, equal to the distance above the convex hull.***

| **Candidate ID** | **s = 0** | **s = 0.25** | **s = 0.50** | **s = 0.75** | **s = 1** |
|---|---|---|---|---|---|
| **A** | +73.78 | +37.44 | +81.76 | +29.06 | +203.37 |
| **B** | +16.45 | +18.52 | +11.04 | +12.76 | +12.97 |
| **C** | +27.71 | +26.99 | +21.10 | +26.94 | +62.18 |
| **D** | +26.56 | +27.99 | +29.45 | +28.33 | +28.33 |
| **G1** | +5.01 | +11.01 | +9.01 | +11.01 | +8.01 |
| **G2** | +3.37 | +5.60 | +4.73 | +6.24 | +4.46 |

Energy of the reaction $NaBH_4 + Ca(BH_4)_2 \rightarrow NaCa(BH_4)_3$, in meV per atom, for the six lattice structures carried to first principles and each cation arrangement s. A positive value places the ternary above the tie-line of the two binary borohydrides, so its decomposition into them is exothermic by that amount at 0 K. Because these two phases span the Na-Ca-B-H convex hull at this composition in the ratio 6:11, the entries are also distances above the hull. Compound and references share the settings of Section 2.3, so all entries are static-lattice energies on one scale.

***Table 5. Converged first-principles energies (eV/atom) of the lowest arrangement at each swap fraction.***

| Candidate ID | s = 0 | s = 0.25 | s = 0.50 | s = 0.75 | s = 1 |
|---|---|---|---|---|---|
| A | −4.115 | −4.152 | −4.107 | −4.160 | −3.985 |
| B | −4.172 | −4.170 | −4.178 | −4.176 | −4.176 |
| C | −4.161 | −4.162 | −4.168 | −4.162 | −4.127 |
| D | −4.162 | −4.161 | −4.160 | −4.161 | −4.160 |
| G1 | −4.184 | −4.178 | −4.180 | −4.178 | −4.181 |
| G2 | −4.186 | −4.183 | −4.184 | −4.183 | −4.184 |

***Table 6. Formation energies per atom (eV) relative to the elemental standard states (BCC Na, FCC Ca, rhombohedral B, molecular $H_2$), for every lattice structure and cation arrangement. Compound and elemental references share identical settings (Section 2.3), so the values are strictly comparable.***

| **Candidate ID** | **s = 0** | **s = 0.25** | **s = 0.50** | **s = 0.75** | **s = 1** |
|---|---|---|---|---|---|
| A | −0.2765 | −0.3128 | −0.2685 | −0.3212 | −0.1469 |
| B | −0.3338 | −0.3317 | −0.3392 | −0.3375 | −0.3373 |
| C | −0.3225 | −0.3233 | −0.3291 | −0.3233 | −0.2881 |
| D | −0.3237 | −0.3223 | −0.3208 | −0.3219 | −0.3219 |
| G1 | -0.3452 | -0.3392 | -0.3412 | -0.3392 | -0.3422 |
| G2 | −0.3469 | −0.3446 | −0.3455 | −0.3440 | −0.3458 |

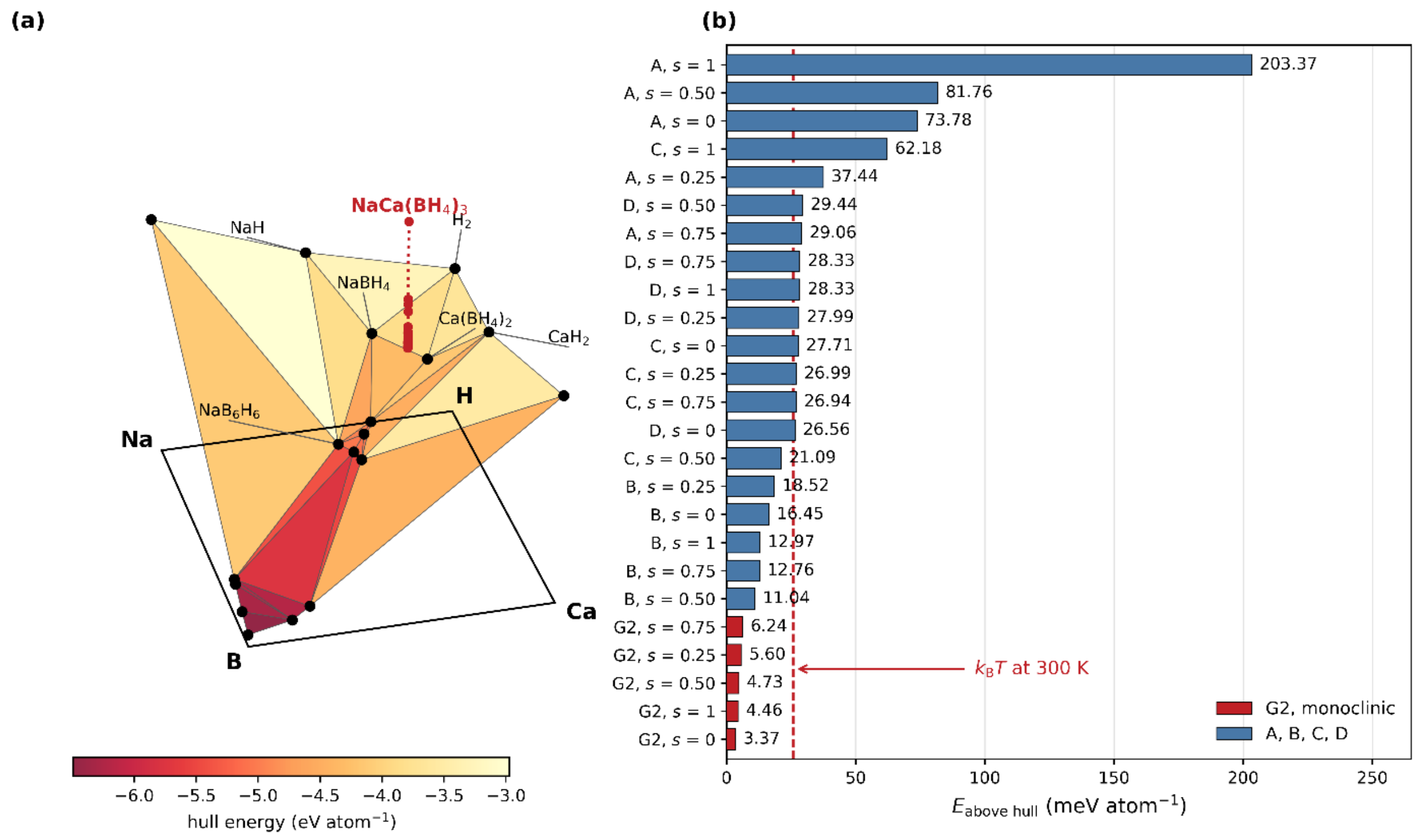

***Figure 5. The $NaCa(BH_4)_3$ candidates on the Na-Ca-B-H convex hull. (a) The hull, built from the 3319 MPDD structures of the system on one machine-learning scale, projected onto a square simplex with corners Na, B, Ca and H and colored by energy. Black points mark the seventeen hull phases, and the candidates do not enter the construction. At $NaCaB_3H_{12}$ the hull is spanned by $NaBH_4$ and $Ca(BH_4)_2$ in the ratio 6:11, so it coincides with the binary tie-line. The red column shows the twenty-five configurations at their first-principles distances of Table 4, magnified fifteenfold against a surface spanning 5.3 eV. (b) The same distances ranked, G2 in red and the four pre-search frameworks in blue. The dashed line marks $k_BT$ at 300 K.***

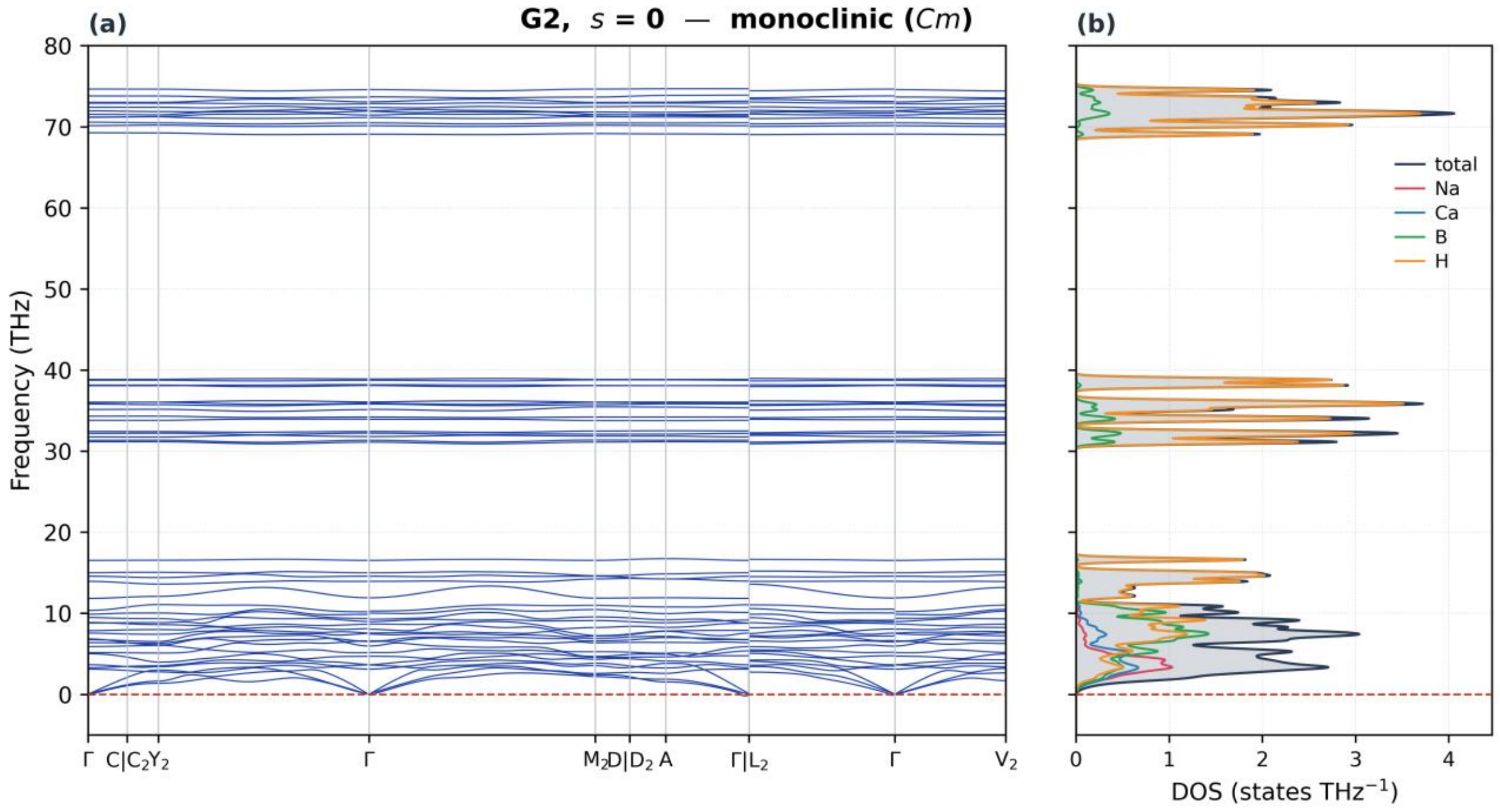

***Figure 6. Lattice dynamics of G2 in its ordered cation arrangement. (a) First-principles phonon dispersion with non-analytical corrections. The twelve wavevectors the supercell resolves exactly are all positive, the lowest at +1.374 THz. (b) Total and atom-projected phonon density of states from the same force constants, convolved with a 0.20 THz Gaussian that removes mesh noise without displacing a band edge. The non-analytical correction is omitted here because it acts only as $q \to 0$ and leaves the density of states unchanged. The cation projections vanish above about 11 THz.***

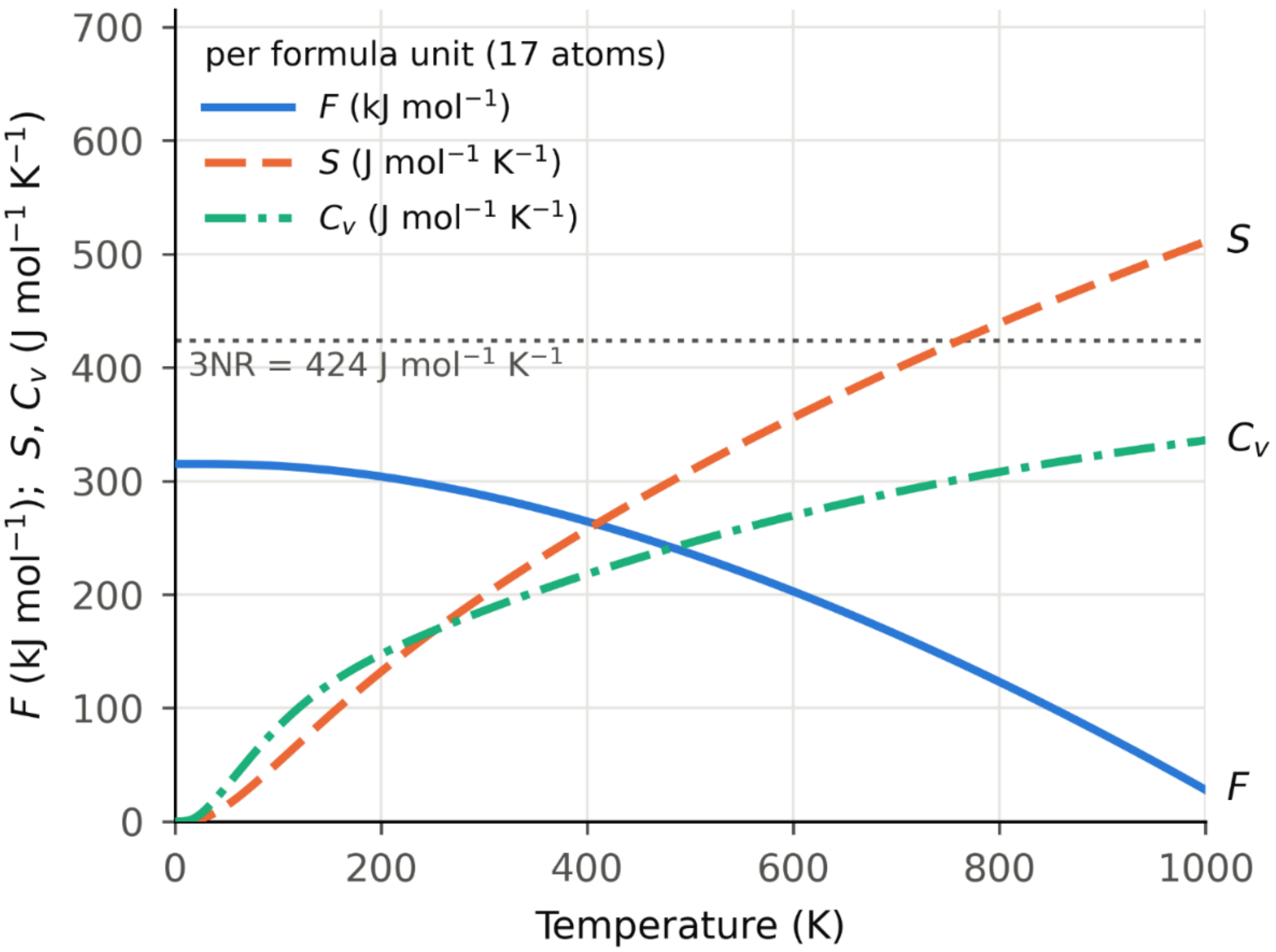

***Figure 7. Harmonic vibrational thermodynamics of G2 per formula unit (17 atoms), from the force constants of Figure 6: Helmholtz free energy F (solid, kJ mol$^{-1}$), vibrational entropy S (dashed) and constant-volume heat capacity $C_v$ (dash-dotted), both in J mol$^{-1}$ K$^{-1}$. F at 0 K is the zero-point energy, 315.0 kJ mol$^{-1}$; the dotted line is the Dulong-Petit limit, 3NR = 424 J mol$^{-1}$ K$^{-1}$.***

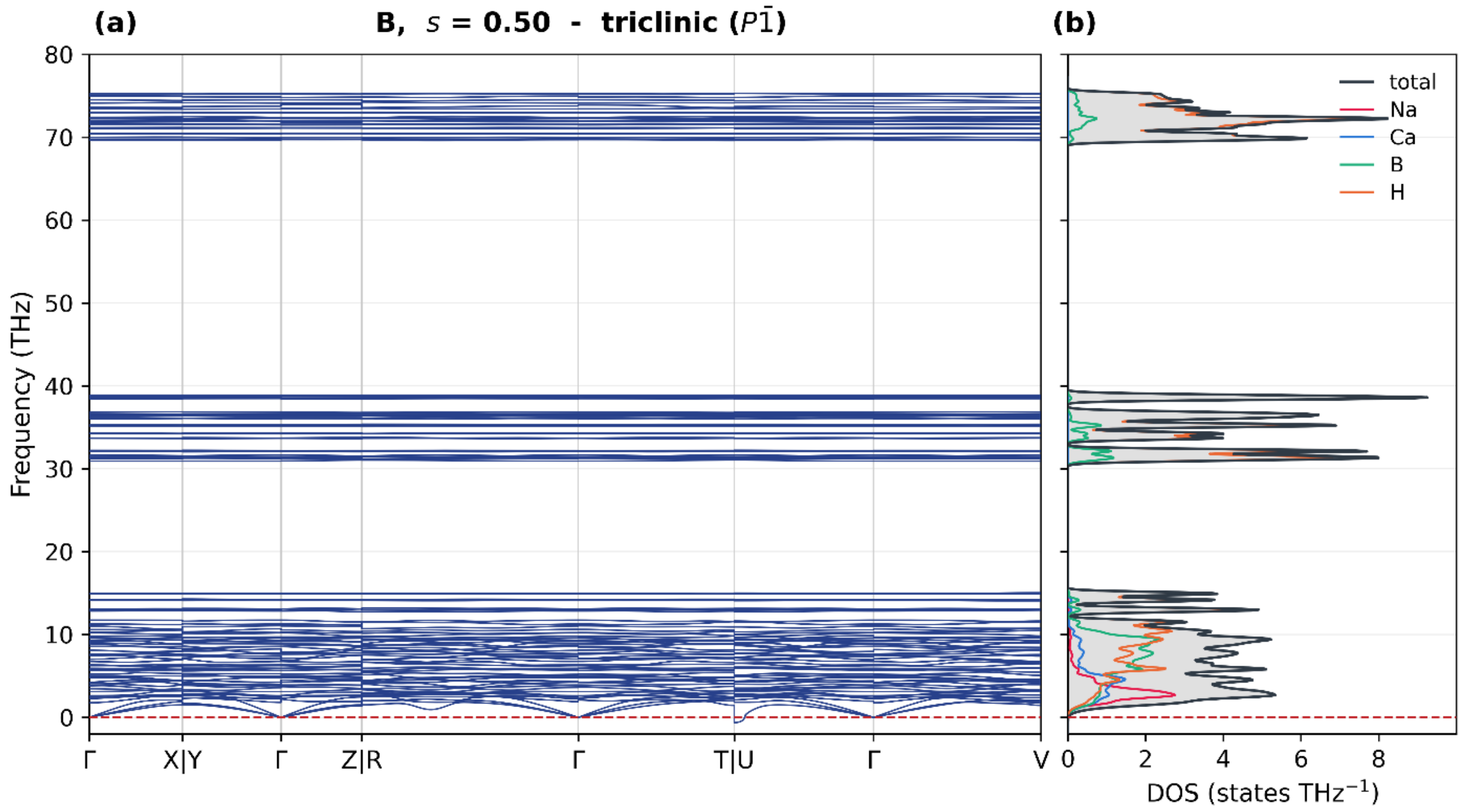

***Figure 8. Lattice dynamics of B at s = 0.50, the most stable perovskite-type lattice structure, from a 2 × 2 × 2 supercell of the 34-atom cell. (a) First-principles phonon dispersion. One imaginary branch reaches −0.635 THz at the zone-boundary wavevector q = (½, 0, ½), which the supercell resolves exactly. (b) Total and atom-projected phonon density of states from the same force constants: external modes below 15.0 THz, H-B-H bending between 30.9 and 38.9 THz and B-H stretching between 69.6 and 75.3 THz.***

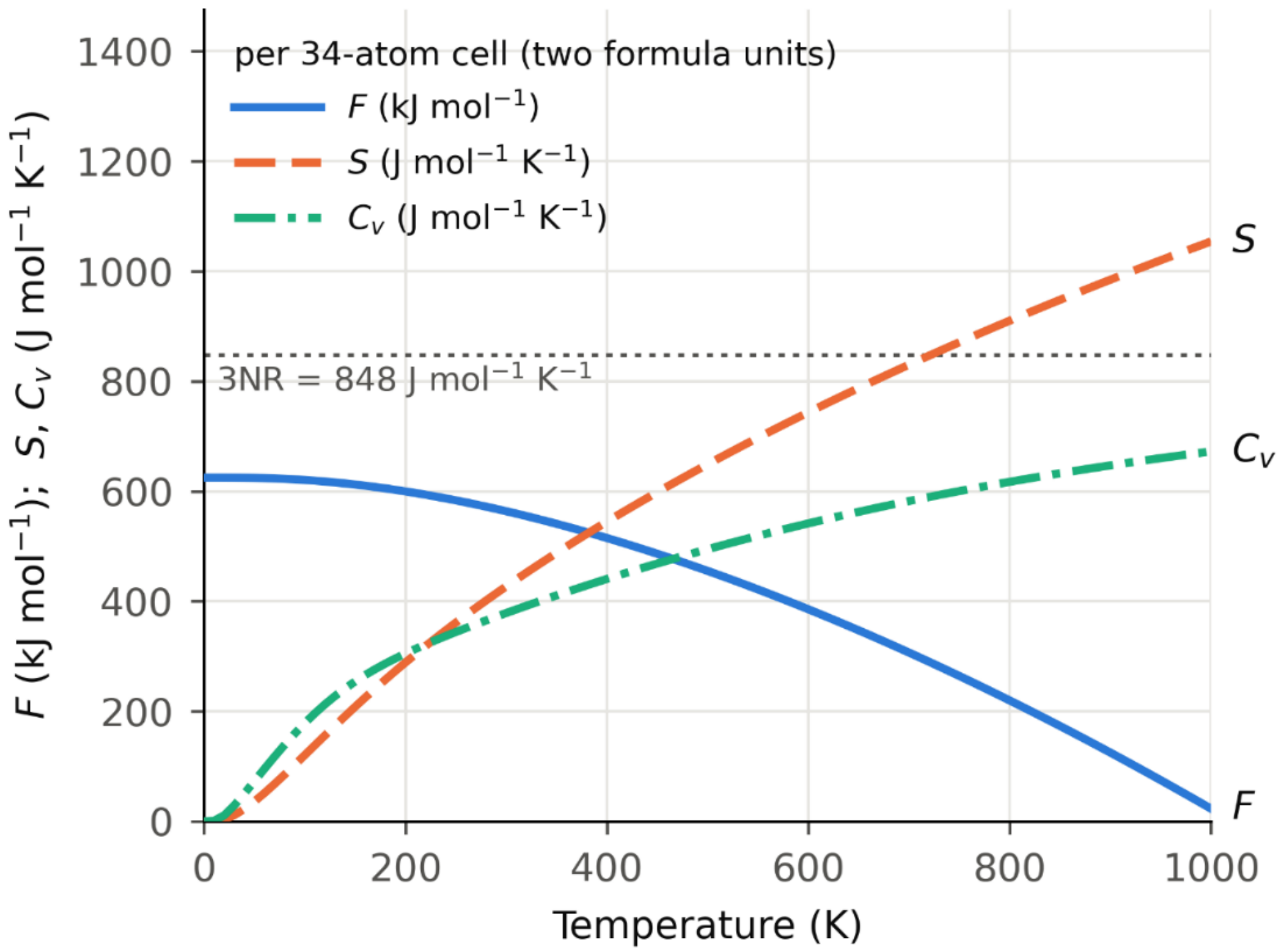

***Figure 9. Harmonic vibrational thermodynamics of B at s = 0.50 per 34-atom cell (two formula units), from the force constants of Figure 8; curves as in Figure 7. F at 0 K is the zero-point energy, 624.9 kJ mol⁻¹; the dotted line is the Dulong-Petit limit, 3NR = 848 J mol⁻¹ K⁻¹.***

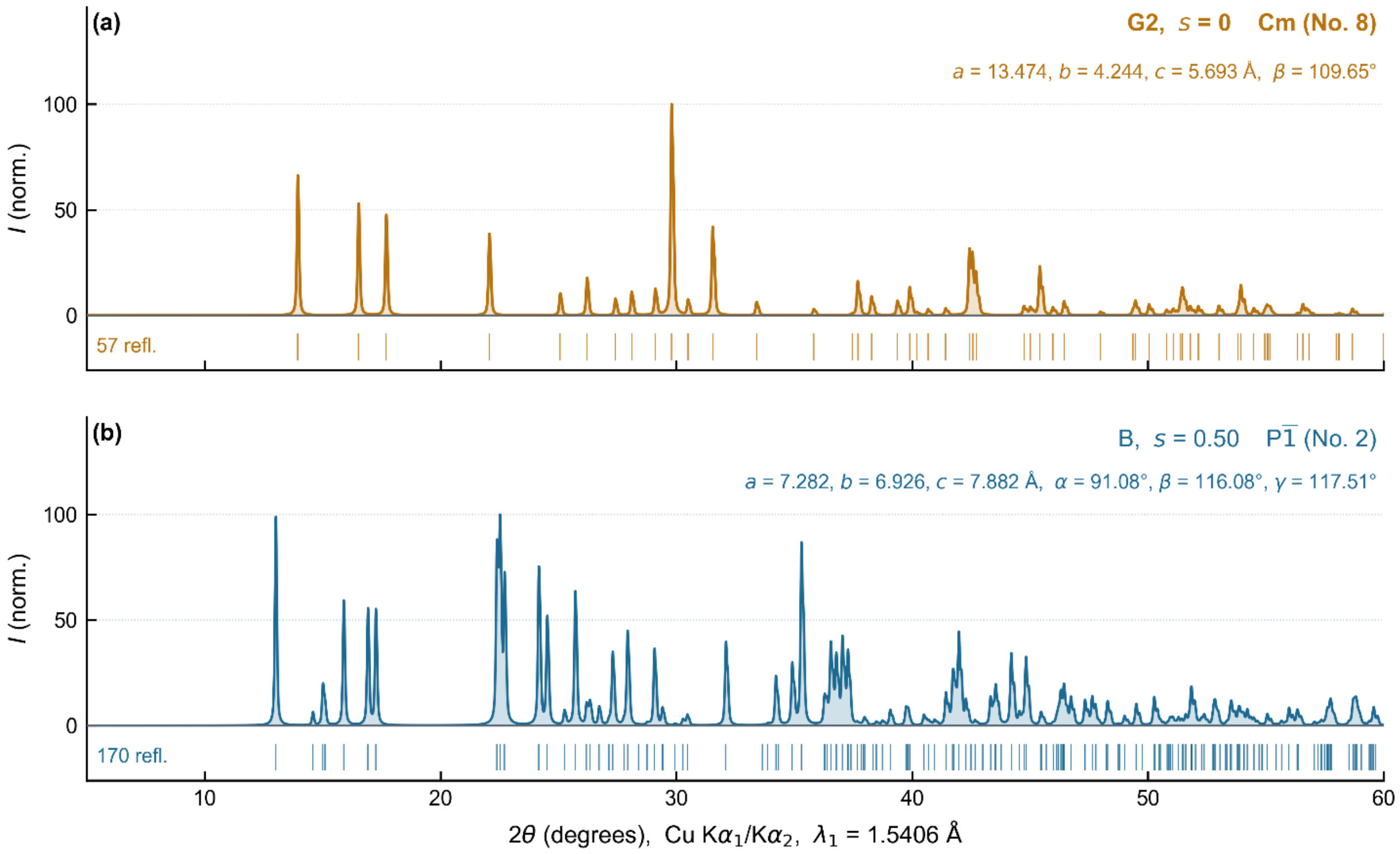

***Figure 10. Simulated powder X-ray diffraction of the most stable lattice structure, G2 at s = 0 (Cm), and of its closest competitor, B at s = 0.50 ($P\overline{1}$), computed at 0 K from the first-principles relaxed cells under the instrument model of Section 2.4. Each pattern is normalized to its strongest reflection. Marks beneath each trace show every symmetry-allowed reflection in the range, 57 for G2 and 170 for B, with the count at the left. The patterns carry no hkl labels because the strong reflections of B lie 0.65° apart on average, against 1.58° for G2, so no labeling fits both panels. The diagnostic reflections of G2 are listed in the text.***